\documentclass{cupconf}

  \checkfont{eurm10}
  \iffontfound
    \IfFileExists{upmath.sty}
      {\typeout{^^JFound AMS Euler Roman fonts on the system,
                   using the 'upmath' package.^^J}%
       \usepackage{upmath}}
      {\typeout{^^JFound AMS Euler Roman fonts on the system, but you
                   dont seem to have the}%
       \typeout{'upmath' package installed. cupconf.cls can take advantage
                 of these fonts,^^Jif you use 'upmath' package.^^J}%
      }
  \else
  \fi

  \checkfont{msam10}
  \iffontfound
    \IfFileExists{amssymb.sty}
      {\typeout{^^JFound AMS Symbol fonts on the system, using the
                'amssymb' package.^^J}%
       \usepackage{amssymb}%
         
       \let\ge=\geqslant  
      }{}
  \fi

  \IfFileExists{amsbsy.sty}
    {\typeout{^^JFound the 'amsbsy' package on the system, using it.^^J}%
     \usepackage{amsbsy}}
    {}

\newsavebox{\astrutbox}
\sbox{\astrutbox}{\rule[-5pt]{0pt}{20pt}}

\title[Initial Mass Function in Clusters]{The Initial Mass Function in Clusters}

\author[B.G. Elmegreen]%
{B\ls R\ls U\ls C\ls E\ns G.\ns E\ls L\ls M\ls E\ls G\ls R\ls E\ls
E\ls N$^1$}

\affiliation{$^1$IBM Research Division, T.J. Watson Research
Center, Yorktown Hts., NY 10598, USA}

\pubyear{2006} \volume{999} \pagerange{1-20}
\date{?? and in revised form ??}
\begin{document}

\maketitle

\begin{abstract}
The stellar initial mass function (IMF) in star clusters is
reviewed. Uncertainties in the observations are emphasized. We
suggest there is a distinct possibility that cluster IMFs vary
systematically with density or pressure. Dense clusters could have
additional formation processes for massive stars that are not
present in low density regions, making the slope of the upper mass
IMF somewhat shallower in clusters. Observations of shallow IMFs
in some super star clusters and in elliptical galaxies are
reviewed. We also review mass segregation and the likelihood that
peculiar IMFs, as in the Arches cluster, result from segregation
and stripping, rather than an intrinsically different IMF. The
theory of the IMF is reviewed in some detail. Several problems
introduced by the lack of a magnetic field in SPH simulations are
discussed. The universality of the IMF in simulations suggests
that something more fundamental than the physical details of a
particular model is at work. Hierarchical fragmentation by any of
a variety of processes may be the dominant cause of the power law
slope. Physical differences from region to region may make a
slight difference in the slope and also appear in the low-mass
turnover point.
\end{abstract}

to be published in "Massive Stars: From Pop III and GRBs to the
Milky Way," eds. M. Livio and Eva Villaver, Cambridge Univ. Press,
in press, from a conference held at the Space Telescope Science
Institute, May 8-11, 2006.

\firstsection 
\section{Introduction: Uncertainties}

The stellar initial mass function (IMF) is difficult to measure
because of systematic uncertainties, selection effects, and
statistical variance. Stars in clusters may all have the same age
and distance, making their masses relatively straightforward to
determine, but mass segregation, field star contamination,
variable extinction, and small number statistics can be problems
in determining the IMF. Nearby clusters show the low mass stars
well, but these clusters also tend to be the most common and
therefore among the lowest in mass, so they do not sample far
enough in the high mass IMF to contain massive stars. High mass
clusters contain massive stars, but these clusters are rare and
the nearest are typically too far away to reveal their lowest mass
stars and brown dwarfs. The Orion trapezium cluster is one of the
few regions where an IMF can be determined throughout all stellar
types, but even then the highest mass star is only one-half or
one-third the mass of the highest possible mass for a star
(Hillenbrand \& Carpenter 2000). No cluster has yet been observed
over the entire stellar mass range. Most of what we know about the
IMF in clusters is from piecing together different parts of the
IMF from different clusters.

Stars in an OB association are also at about the same distance
from the Sun, but they typically span a range of ages that is
longer than the shortest lifetime of a massive star, making
formation rate and stellar evolution corrections necessary before
determining the IMF. OB associations also tend to have variable
extinction, and their dispersal has to be considered to
reconstruct which stars actually formed there.

The IMF in the field comes partly from stars that formed in the
field (in small molecular clouds), partly from stars that drifted
there out of nearby OB associations, and partly from old dissolved
clusters. The advantage of IMF determinations in the field is that
tens of thousands of stars can be included in a survey (e.g. ,
Parker et al. 1998), as can a wide range of stellar masses.
However there are many uncertainties in converting what is
observed, the present day mass function, into what is desired, the
initial mass function. For example, this conversion depends on the
star formation history and the rate of vertical disk heating.
Stellar evolution and the mass-luminosity relation are also
important.

The IMF in whole galaxies comes from the summed IMFs of all the
star forming regions, i.e., from the clusters, loose groups,
associations and even the accreted satellite stars.  Average IMFs
are typically derived from abundance ratios (e.g., iron comes
mostly from low mass stars and oxygen comes mostly from high mass
stars), color magnitude diagrams, H$\alpha$ equivalent widths,
etc.. However, resolution limits, faintness, unknown star
formation histories, variable extinction, crowding, and many other
problems can arise in the determination of a galaxy-wide IMF.

\section{IMFs in clusters: should we expect
systematic variations?}

Many dense clusters have an IMF with a slope at intermediate to
high mass that is close to the Salpeter IMF slope, $\Gamma=1.35$
on a plot with log-mass intervals ($\xi(M)d\log M\propto
M^{-\Gamma}d\log M$), which is the same as a negative slope of
$1+\Gamma$ on a plot with linear intervals in mass. There is
considerable variation around this slope ($\pm0.5$ in Scalo 1998),
but this could be from sampling statistics (Elmegreen 1999; Kroupa
2001)). The 30 Dor cluster has a slope remarkably close to the
Salpeter value (Massey \& Hunter 1998), as do the clusters h and
$\chi$ Persei (Slesnick, Hillenbrand \& Massey 2002), NGC 604 in
M33 (Gonz\'alez Delgado \& Perez 2000), NGC 1960 and NGC 2194
(Sanner et al. 2000), NGC 6611 (Belikov et al. 2000), and many
others.

The Sco-Cen OB association has an IMF significantly steeper than
the Salpeter function (Preibisch et al. 2002); the slope is $-1.7$
to $-1.8$ instead of $-1.35$. The massive stars in W51 ($M\ge4$
M$_\odot$) also have a steep IMF slope, $-1.8$, but two of the
four subgroups in this region have a statistically significant
excess of stars at the highest mass ($\sim60$ M$_\odot$; Okumura
et al. 2000). We cannot tell whether these are physical variations
or statistical fluctuations.  Peretto, Andre \& Belloche (2006)
suggested that three cores in the massive star-forming region NGC
2264 appear to be headed for a merger. If some massive stars form
by mergers or other peculiar events, then fluctuations at the high
mass end of the IMF can be large considering that most clusters
form only a few high mass stars anyway.

Indeed, if there are two routes to forming a high mass star, then
there have to be two IMFs, one for the regions that favor one
process and another for the regions that favor the second process.
Two such processes could, for example, include gas core
contraction in a turbulent medium (Tan \& McKee 2004; Krumholz,
McKee, \& Klein 2005), and gas accretion from an intercore medium
(Bonnell et al. 2006). Another process could be protostar
coalescence in a cluster core. The first of these processes has
the stellar mass defined by core formation rather than accretion
or coalescence after core formation, and this first process may
apply to a wide range of environments, including -- but not
limited to -- dense clusters. The second and third of these
processes may work best in dense clusters. If this is the case,
then the cluster IMF would have more routes to the formation of
massive stars than other regions, and could therefore have a
systematically flatter slope. This does not mean it would have an
IMF flatter than the Salpeter IMF, because the turbulent core
process could have a high mass IMF that is steeper than Salpeter.
In that case, all the additional processes that work in a cluster
may serve only to flatten the cluster IMF to the Salpeter slope,
with only the most extreme cluster conditions flattening it more
than the Salpeter slope. Evidence for independent variations at
both the high and low mass ends of the IMF were summarized by
Elmegreen (2004). The observations seem to support the view that
low density regions have slightly steeper IMFs than the Salpeter
slope.  This is consistent with the existence of multiple routes
to massive stars.

\section{Do massive stars need the cluster environment?}

Testi, Palla \& Natta (1999) suggested that Herbig Ae/Be stars
have a correlation between maximum mass and the surrounding
cluster density. The observation was really that more massive
clusters have a more massive upper end of the IMF, but because all
of their clusters have about the same radius, the cluster mass
translates into a cluster density. Bonnell \& Clarke (1999) showed
their result could be from sampling statistics: more massive
clusters sample further out in the IMF. A similar debate took
place 16 years earlier (Larson 1982; Elmegreen 1983). There have
been several other attempts to correlate cluster mass with maximum
stellar mass too (e.g., Khersonsky 1997). The size-of-sample
effect is strong, however, and it can disguise a physical link
between cloud mass and stellar mass making the physical effect
difficult to demonstrate. At the moment, there are no clear
correlations between maximum stellar mass and the cloud or cluster
mass that are in excess of expectations from sampling statistics.

The important question is whether the cluster environment affects
the final stellar mass distribution. We mentioned how it might in
the previous section (denser regions flatten the IMF), but know of
no clear evidence for it one way or another. de Wit, et al (2005)
turned the question around and investigated whether massive stars
ever form alone in the field. They observed 43 local ``field''
O-type stars and looked for evidence that they escaped from a
cluster where they might have formed. Most of these O stars could
reasonably be placed with some nearby cluster, but a few, 4\%
overall, could have formed in isolation. de Wit et al. pointed out
that this percentage is consistent with a cluster mass function
that extends down to a single $\sim100$ M$_\odot$ star with slope
of $\beta=1.7$.

Oey, King \& Parker (2004) did a similar study in the LMC, finding
the distribution function of the number of O-type stars in
clusters. This distribution also went smoothly down to clusters
containing a single O star. The difference between the Oey et al.
result and the de Wit et al. result is that the Oey et al.
clusters also contain other stars, but the local isolated O stars
do not occur in clusters and are truly isolated.  If these
isolated O stars cannot be traced to clusters, and if they really
formed alone or in a loose group, then it would appear that
massive stars do not need the cluster environment. It would be
very interesting to know the IMF of stars which do not form in
dense clusters. The above discussion suggests that this
``isolated'' IMF (not to be confused with a ``field'' IMF, which
is a blend) should be steeper than the cluster IMF.

\begin{figure}
\vspace{4.in} \includegraphics{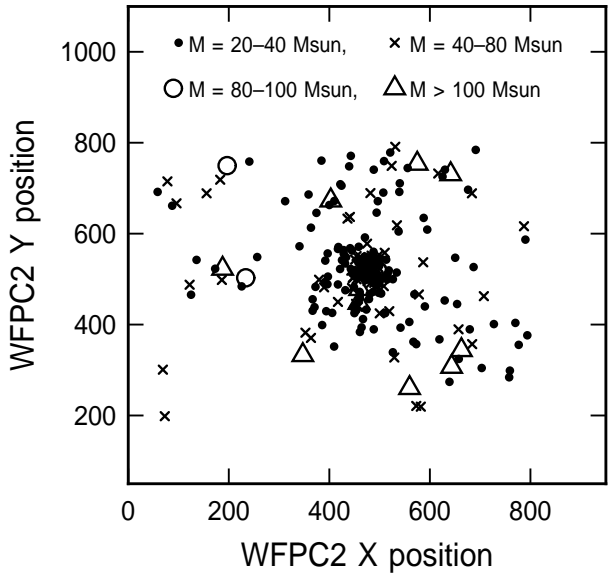}\caption{Distribution of O-type stars in the 30 Dor
region, using data from Massey \& Hunter (1998). High mass stars
form all over the region, not just in the dense
core.}\label{hunterxym}
\end{figure}

Figure \ref{hunterxym} shows the positions of all massive stars in
the 30 Doradus region of the LMC, using data from Massey \& Hunter
(1998).  The various symbols represent stars in different mass
intervals. Clearly the high mass stars appear all over the region,
even outside the dense cluster core, which is R136. This
distribution is not surprising because the peripheral gas is still
dense and fragmented, and it is also compressed by the stars in
the core, leading to triggering (Walborn et al. 1999). Continuing
the discussion of the previous paragraph, it would be interesting
to know the IMF of triggered star formation.

There are many young regions that recently formed O-type stars but
show no evidence for clusters at all. NGC 604 in the galaxy M33 is
an example. Hunter et al. (1996) estimated the massive star IMF
there and derived a slope of $-1.6$. These regions have been
called super OB associations by Ma\'iz-Apell\'aniz (2001), who
studied other examples.  One would think if the O stars formed in
clusters, there would be some remnant or core of those cluster
remaining during the short massive-star lifetime. The lack of such
cores implies the O stars formed in relative isolation.

\section{Applications to Starbursts and Young Elliptical
Galaxies}

Starbursts and mergers have a large fraction of their young stars
in clusters (Larsen \& Richtler 2000), and because the total
number of clusters can be large, the samples can include rare
supermassive clusters (Whitmore 2003).  It is unknown whether star
formation at a high rate in bursts produces the same range of
cluster masses and the same IMF as star formation at a lower rate
for a longer time, both producing the same total mass in stars. If
so, then the cluster mass function and the IMF are sampling from
universal functions. But this need not be the case.  Starbursts
could produce more massive clusters in a short time than normal
galaxies in a long time because the pressure is always higher in a
starburst and massive clusters are high-pressure regions.  If also
follows that if the IMF depends on the cluster environment, or if
there are two IMFs, one of which depends more on the cluster
environment than the other, then starburst regions with a
significant population of massive or unusually dense clusters
could produce a flatter IMF than normal galaxies.

Whole starburst regions do not appear to have IMFs noticeably
flatter than the Salpeter function (see review in Elmegreen 2005),
but some regions may have individual clusters with top-heavy or
bottom-light IMFs. Sternberg (1998) found a high light-to-mass
ratio in NGC 1705-1 that implies either an IMF slope shallower
than $\Gamma=1$ or an inner cutoff to the IMF that removes low
mass stars.  Smith \& Gallagher (2001) found the same for the
cluster M82F; an inner cutoff around 2 to 3 M$_\odot$ for
$\Gamma=1.3$. Alonso-Herrero et al. (2001) derived a high L/M
ratio for clusters in the starburst galaxy NGC 1614.  McCrady et
al. (2003) observed another cluster in M82, MGG-11, and inferred a
deficit in low mass stars.  Mengel et al. (2002) had the same
conclusion for clusters in NGC 4038/9.

Other super star cluster appear to have normal IMFs, so the dense
cluster environment alone does not guarantee a flat IMF. Examples
of normal IMFs are in NGC 1569-A (Ho \& Filippenko 1996; Sternberg
1998), NGC 6946 (Larsen et al. 2001), and MGG-9 in M82 (McCrady et
al. 2003).

Finding the light-to-mass ratio in a cluster is a difficult
problem. The velocity dispersion in the cluster has to be measured
along with the radius to get the mass, and the luminosity has to
be measured. However, the velocity dispersion could vary with
radius inside a young cluster, in which case the observed
dispersion is a weighted integral over the position, and then the
isothermal expression would not apply to the conversion between
velocity dispersion and radius. The proper radius to use is
uncertain because the cluster could be evaporating, out of
equilibrium, non-isothermal, multi-component, non-isotropic, or
non-centralized (Bastian \& Goodwin 2006). Also, the core could
poorly resolved. The choice of aperture for the velocity
dispersion measurement is difficult. Field star corrections may be
necessary for both the density profile and the velocity
dispersion.

The IMFs of massive elliptical galaxies appear to be slightly
flatter than in spiral galaxies (Pipino \& Matteucci 2004;
Nagashima et al. 2005b). Clusters of galaxies also suggest a
history of top-heavy IMFs in the form of elliptical galaxy
starbursts (Renzini et al. 1993; Loewenstein \& Mushotsky 1996;
Chiosi 2000; Moretti, Portinari, \& Chiosi 2003; Tornatore et al.
2004; Romeo et al. 2005; Portinari et al. 2004; Nagashima et al.
2005a). Low surface brightness gal. may have steeper-than-normal
IMFs (Lee et al. 2004).

Taken together, these observations suggest a possible excess of
high mass stars in some starburst clusters or in early phase
starburst elliptical galaxies, and a possible deficit of high mass
stars in the most quiescent environments (low surface brightness
galaxies). This trend is consistent with the existence of several
routes to the formation of a high mass star, with at least one of
these routes more active in the type of environment that has a
high star formation rate. The exact physical processes that are
involved with this ``starburst route'' are not observed yet, but
enhance accretion, protostellar coalescence, and high thermal
temperatures would all work in this direction.

\section{Mass segregation}

A problem with IMF determinations for clusters is mass
segregation, where the most massive stars are either born near the
center or migrate toward the center after a random walk of
scattering events. The IMF is often observed to be shallower in
the central regions of clusters.  The nearest large cluster, NGC
3603, has a relatively shallow IMF slope in the core and a
relatively steep IMF slope near the edge (Sung \& Bessell 2004)
There is possibly a high mass drop-off in this cluster too: the
slope is $-1.9$ overall for $M>40$ M$_\odot$ (steeper than the
Salpeter slope, which is $-1.35$).  The Orion cluster has mass
segregation too, prompting Hillenbrand \& Hartmann (1998) to
suggest it was there from birth, considering the young age of this
cluster. At even younger age, the mm-wave continuum sources in
Ophiuchus appear mass segregated (Elmegreen \& Krakowski 2001).

A good example of a flat IMF that could either be top-heavy as in
some super star clusters, or mass-segregated, is in the Arches
cluster. Yang et al. (2002) and Stolte et al. (2005) found an IMF
slope of $\Gamma\sim -0.8$ there.  The Arches cluster is in a
region of intense tidal forcing (the galactic center) and it seems
plausible that the outer, low-density regions have been tidally
stripped, leaving only the dense core. If the dense core was as
mass segregated as other clusters, which show the same flat slopes
in their cores (e.g., de Grijs, et al. 2002), then Arches would
not have an unusual IMF.

The case of tidal stripping for the Arches cluster seems
compelling after similarly flat IMFs have been observed in tidally
stripped halo globular clusters. de Marchi, Pulone, \& Paresce
(2006) show a flat mass function in the galactic cluster NGC 6218.
At four different radii, the mass function slopes in their figure
are $+1.4$, $+1.3$, $+0.6$, and $+0.1$ (note the positive values,
when the Salpeter slope on a comparable figure is $-2.3$). Flat
mass functions are also seen in the globular clusters NGC 6712 (de
Marchi et al. 1999) and Pal 5 (Koch et al. 2004). These latter two
are expected to have undergone tidal stripping. Tidal stripping is
suspected in NGC 6218 as well; the observed cluster mass is
supposed to be only 20\% of the original mass. Recent models by
Baumgardt (2006) show how tidal stripping can leave a cluster with
a flat IMF.

\section{The Low mass IMF in clusters}

The low mass part of the IMF has been observed down to and beyond
the brown dwarf regime in nearby young clusters where brown dwarfs
are still bright on their pre-main sequence tracks. The count of
low mass stars is usually fairly high, making statistical
fluctuations in the IMF much smaller than at higher masses.
Generally the IMF turns over from its $\Gamma\sim1.35$ type slope
at intermediate to high mass and becomes somewhat flat with
$\Gamma\sim0$. The turnover occurs somewhere between 0.1 M$_\odot$
and 1 M$_\odot$ (Scalo 1986; Kroupa 2001; Chabrier 2003).

Several authors have noted variations in the relative abundance of
low and high mass stars, or in the shape and extent of the flat
low mass part. For example, IC 348 (Preibisch, Stanke \& Zinnecker
2003; Muench et al. 2003; Luhman et al. 2003) and Taurus (Luhman
2000; Brice\~no et al. 2002) have brown dwarf-to-star ratios that
are $\sim2$ times lower than the Orion trapezium cluster
(Hillenbrand \& Carpenter 2000; Luhman et al. 2000; Muench et al.
2002), Pleiades (Bouvier et al. 1998; Luhman et al. 2000), M35
(Barrado y Navascu\'es et al. 2001), and the galactic field (Reid
et al. 1999).  There are many possible reasons for such
fluctuations, but no observations yet favor one reason over
another.

\section{Theory}

Stars form in dense molecular cores where self-gravity overcomes
magnetic and pressure forces and where turbulent motions are too
slow to disrupt the gas in a free fall time. The origin of the
cores is not fully understood. They are likely to have several
formation mechanisms, including compression in shocks caused by
turbulence and stellar outflows.  Protostars form in the cores,
but what happens after that is also unclear. If the protostars are
rapidly converging in a cloud-wide collapse, then they can move
together and interact strongly, creating tightly bound systems
that disperse quickly (Bonnell et al. 2001). If the cores and
protostars move slowly or have a low space density, then they will
not interact. Strong interactions could make the stellar mass
function different from the core mass function, while weak
interactions might keep them about the same.  This duality led to
the suggestion above that there could be two high-mass IMFs
appearing in different regions, depending on the degree of core
and protostellar interactions.

Most simulations get the observed IMF, but then there are enough
tunable parameters to assure this result if it is desired. Recent
simulations have been probing the sensitivity of the results to
the assumptions. Bate \& Bonnell (2005) did two SPH simulations
with no magnetic field, each having a different initial thermal
Jean mass. They found that the characteristic or turnover mass in
the resulting IMF scaled directly with the input Jeans mass.
Jappsen et al. (2005) considered variations in the equation of
state. If the ratio of specific heats or adiabatic index $\gamma$
is less than 1 at low density and greater than 1 at high density,
meaning that the equation of state gets stiffer at high density,
then the mean mass in the resulting IMF is comparable to the Jean
mass at this transition density (see also Larson 2005). If the
transition density is higher, the Jeans mass becomes lower and
there are more cores resulting. Martel et al. (2006) did SPH
simulations with particle splitting, no magnetic fields and an
isothermal equation of state. They found that the characteristic
mass in the resulting IMF depended on resolution: when the number
of levels in the splitting hierarchy increased, and lower masses
could be resolved, the mean core mass decreased in proportion.

Li et al. (2004) did magnetohydrodynamic (MHD) super-Alfvenic
simulations on grids of various sizes and got a power law IMF with
a turn over at low mass, in reasonable agreement with
observations. Tilley \& Pudritz (2005) did MHD simulations in a
$256^3$ grid, testing the implications of different ratio of
gravitational to magnetic energies. Their preferred model for the
IMF had comparable thermal and magnetic pressures, and very large
ratios of gravitational to magnetic energy density -- on the order
of 100.  Padoan et al. (2005) did an adaptive mesh run without
magnetic fields and found that brown dwarfs could form by
turbulent fragmentation, even though their masses were much less
than the initial Jeans mass.  Nakamura \& Li (2005) did a 2D MHD
simulation and showed that magnetic diffusion is enhanced by
turbulent compression; still, the slowness of magnetic diffusion
lowers the efficiency of star formation compared to non-magnetic
simulations.

Several important physical effects are missing from these models.
Feedback that erodes disks and pre-collapse objects is usually not
considered, although Li \& Nakamura (2006) included protostellar
winds as a source of turbulence in their 3D MHD simulations. Also
missing are fully turbulent systems before star formation begins
and anything like a turbulent environment that can affect the
simulation boundary (i.e., variable total grid mass, variable
center of gravity, etc.). Heating and cooling are usually
considered with only crude approximations, like polytropic
assumptions or constant temperatures. Generally only small numbers
of stars form so the modeled IMF is statistically inaccurate. In
the case of SPH models, magnetic forces are always missing. For
MHD on a single grid, there is a limited dynamic range for
density.  MHD simulations also do not treat the physics of
detachment of the background magnetic field from stars.
Nevertheless, it is only a matter of time before these limitations
are overcome and simulations make the IMF in a realistic way.

The lack of magnetic fields in SPH simulations can severely affect
the interpretation of the results.  If the clump magnetic field is
critical, or if the clump forms with a constant mass-to-flux ratio
in a cloud where the average magnetic field is critical, then the
field strength in the clump satisfies, $B_{clump}\sim G^{1/2}
\Sigma_{clump}$ for clump mass column density $\Sigma_{clump}$.
The magnetic force per unit volume acting on the clump as it drags
the field around is $\sim B_{clump}^2/R_{clump}\sim
G\Sigma_{clump}^2/R_{clump}$. The gravitational force per unit
volume acting on the clump by the rest of the cloud is $\sim
G\Sigma_{cloud}\times M_{clump}/R_{clump}^3 =
G\Sigma_{cloud}\Sigma_{clump}/R_{clump}$. Thus the ratio of the
magnetic to gravitational force on a clump that drags the field
around as the clump responds to the gravitational force of the
surrounding cloud is
\begin{equation} F_B/F_G \sim \Sigma_{clump}/\Sigma_{cloud}
>>1.\end{equation} Thus, clumps do not free fall in the cloud
until either their magnetic field lines are detached or their
fields diffuse out. Magnetic field-free models that produce swarms
of freely moving clumps and eventually protostars would seem to be
unrealistic unless the whole clouds are collapsing too.

Similarly, we can calculate the likelihood of clump accretion from
remote parts of the cloud. Magnetic fields should severely limit
such accretion. The magnetic force per unit volume exerted on the
ambient gas in a cloud is $\sim B_{cloud}^2/R_{cloud}\sim
G\Sigma_{cloud}^2 / R_{cloud}$. The gravitational force per unit
volume on this ambient gas that is exerted by the clump is $\sim
(GM_{clump}/R_{cloud}^2)\times M_{cloud}/R_{cloud}^3$. The
magnetic to gravitational force ratio for accreted ambient cloud
gas is
\begin{equation} F_B/F_G \sim
M_{cloud}/M_{clump}>>1.\end{equation} Thus the ambient cloud gas
cannot freely fall onto a single clump or even onto a cloud core
whose mass is significantly less than the mass of the whole cloud.
These two effects from magnetic fields in clouds would seem to be
a problem for magnetic-field free simulations where gravitational
and turbulent motions turn into freely moving protostars which
competitively accrete gas from the whole cloud. More likely,
protostars accrete from their immediate neighborhoods or clumps,
and they do not move rapidly in a cloud until their field lines
are almost completely detached from the background.

There are also other magnetic effects that are not present in SPH
simulations and which could be important for cloud structure and
star formation. One large issue concerns dynamical communication
between the cloud and the surrounding ISM. Magnetic fields connect
the cloud, the cloud cores, and all of the pre-detached clumps to
the external ISM. Magnetic stresses thereby transfer linear and
angular momentum from inside the cloud to outside, and vice versa.
These stresses are a source of damping for clump and cloud
turbulent motions, and a possible source of energy into these
motions from external turbulence (Elmegreen 1981).  Internal
feedback should be more influential on cloud structure when the
cloud magnetic field is near the energy equilibrium value (the
critical field). If the field is subcritical, it may have a more
limited role (Padoan \& Nordlund 1999).

\section{Reflections}

Computer simulations usually make IMFs like the observed IMF, but
we should question whether they do this for the right reasons
because all of the different models have different assumptions and
physical processes at work. The universality of the real IMF
suggests an insensitivity to detailed processes. The IMF is the
same inside and outside of star clusters, and it is about the same
for starbursts and for slow star formation in galaxies. It is also
nearly independent of metallicity, galaxy mass, and epoch in the
Universe after some heavy elements form.  With similar
insensitivities, the simulations could also get the right result
even if the assumed physics were oversimplified.

For example, hierarchical fragmentation alone gives a mass
function $n(M)dM\sim M^{-2}dM$, which is very close to the
Salpeter IMF ($M^{-2.35}dM$).  What if the modeled IMF came mostly
from fragmentation, regardless of the origin of this
fragmentation? Such an IMF would be mostly the result of geometric
effects. Then, if physical processes make star formation slightly
more likely at intermediate mass (i.e., at about the Jeans mass
$M_J$), the observed IMF could follow. That is, physical processes
favoring $M_J$, in addition to fragmentation, could steepen the
IMF from a function like $M^{-2}$ to the observed function
$M^{-2.4}$ for $M>M_J$. Similarly, a bias toward $M_J$ would
flatten the pure-fragmentation IMF from $M^{-2}$ to $M^{-1.5}$ for
$M<M_J$. Additional processes could act in dense clusters to make
an excess of massive stars, or an excess of brown dwarfs. These
additional processes might include the ablation of low mass
protostars, heightened accretion, coalescence, and multiple star
interactions.

\section{Conclusions}

Observations suggest a more or less constant IMF in many diverse
environments. There are hints of variations at the high and low
mass ends of the IMF, suggesting perhaps a tri-modal IMF. Many
things could cause these variations, such as protostellar
coalescence, enhanced gas accretion, multiple system star
ejections, and so on, as discussed extensively in the literature.
There are also possible false variations of the IMF from unknown
star formation histories, incorrect mass-to-light ratios, field
star contaminations, small number statistics, and so on.

The theory of gravo-turbulent fragmentation typically gets the
observed IMF, but many uncertainties remain. Magnetic fields,
feedback, boundary conditions and initial conditions are all
concerns. Yet the diverse models usually get about the right IMF.
If the simulations get the right IMF even under highly simplified
conditions, it is fair to ponder what the simulations and reality
have in common that always gives this IMF. Perhaps it is
fragmentation alone, whether from gravity or turbulence and
independent of the proportion. Perhaps it is from accretion alone,
as suggested by Bonnell et al. (2006). It could even be from the
large number of independent parameters in both simulations and
reality, which through random variations, bring the system to a
common mass function that is log-normal. More simulations are
needed before these questions can be clarified.

\end{document}